\documentclass{kluwer}    
\usepackage{graphicx,psfig}

\newdisplay{guess}{Conjecture}

\begin{document}  
\begin{article}
\begin{opening}         
\title{The Mass to Light ratio and the Initial Mass Function in galactic discs}
\author{Laura \surname{Portinari}}  
\author{Jesper \surname{Sommer--Larsen}}  
\institute{Theoretical Astrophysics Center, Copenhagen, Denmark}
\author{Rosaria \surname{Tantalo}}  
\institute{Dipartimento di Astronomia, Padova, Italy}
\runningauthor{L.~Portinari et~al.}
\runningtitle{The M/L ratio and the IMF in galactic discs}

\begin{abstract}
A low mass--to--light ratio for the baryonic component 
of spiral galaxies is advocated by a number of dynamical 
studies and by cosmological simulations of galaxy formation.
We discuss the possibility of obtaining low mass--to--light ratios 
for the stellar component in discs, by changing the Initial Mass Function
and the Star Formation History.
\end{abstract}
\end{opening}           
\vspace{-3mm}

\section{Introduction}  

\vspace{-3mm}
\noindent
Recent N--body+SPH cosmological simulations of the formation of disc galaxies 
reproduce the observed Tully-Fisher (TF) relation,
provided the mass--to--light (M/L) ratio of the stellar component is 
low, {\mbox{M/L$_I \sim$0.8}} in the I--band 
(Sommer-Larsen \& Dolgov 2001, Sommer-Larsen et~al.\ 2002). 
Various arguments support in fact this possibility.

A straightforward estimate of the stellar mass of the Milky Way yields
{\mbox{M$_* \sim 5 \times 10^{10}~M_{\odot}$}}; 
to lie on the observed TF relation as other spirals,
its M/L$_I$ must be $\sim$0.8 (Sommer-Larsen \& Dolgov 2001).

Bar instability arguments put upper limits on the
M/L ratio of discs, M/L$_I \leq 1.85~h$, i.e.\
M/L$_I \leq 1.2$ for $h=0.65$ (Syer et~al.\ 1997).

Even in case of maximal stellar discs, lower M/L ratios
for the stellar component are required, than those predicted by
the Salpeter IMF (Bell \& de Jong 2001). And it is still much debated
whether discs are maximal or sub--maximal;
for his favoured sub--maximal disc model, Bottema (2002) finds
M/L$_I \sim 0.82$.

Finally, two recent dynamical studies of individual spiral galaxies yield 
{\mbox{M/L$\sim$1}}
in the B, V and I band for the Sc galaxy NGC 4414 (Vallejo et~al.\ 2002)
and M/L$_I$=1.1 for the disc of the Sab spiral 2237+0305, 
Huchra's lens (Trott \& Webster 2002).

\vspace{-3mm}

\section{Initial Mass Function and Star Formation History }

\vspace{-3mm}

\noindent
A standard Salpeter IMF, extended over the mass range [0.1--100]~$M_{\odot}$, 
certainly yields much higher M/L ratios than those mentioned above
(Fig.~2, top left). 
There is however plenty of observational evidence that the IMF
flattens below $\sim 1 M_{\odot}$, possibly 
with a turn--over at low masses, and hence is ``bottom--light'' with respect
to the Salpeter IMF.

Here we consider the following IMFs (Fig.~1, left panel): 
the {\bf Salpeter} IMF;
the {\bf Kroupa} (1998) IMF, derived from field stars in the Solar 
Neighbourhood;
the {\bf Kennicutt} IMF, derived from the global 
properties of spiral galaxies (Kennicutt et~al.\ 1994);
the {\bf Larson} (1998) IMF, with an exponential cut--off 
at low masses as favoured
by recent determinations of the local IMF down to the sub--dwarf regime
(Chabrier 2001, 2002).

\begin{figure}
\leavevmode
\psfig{file=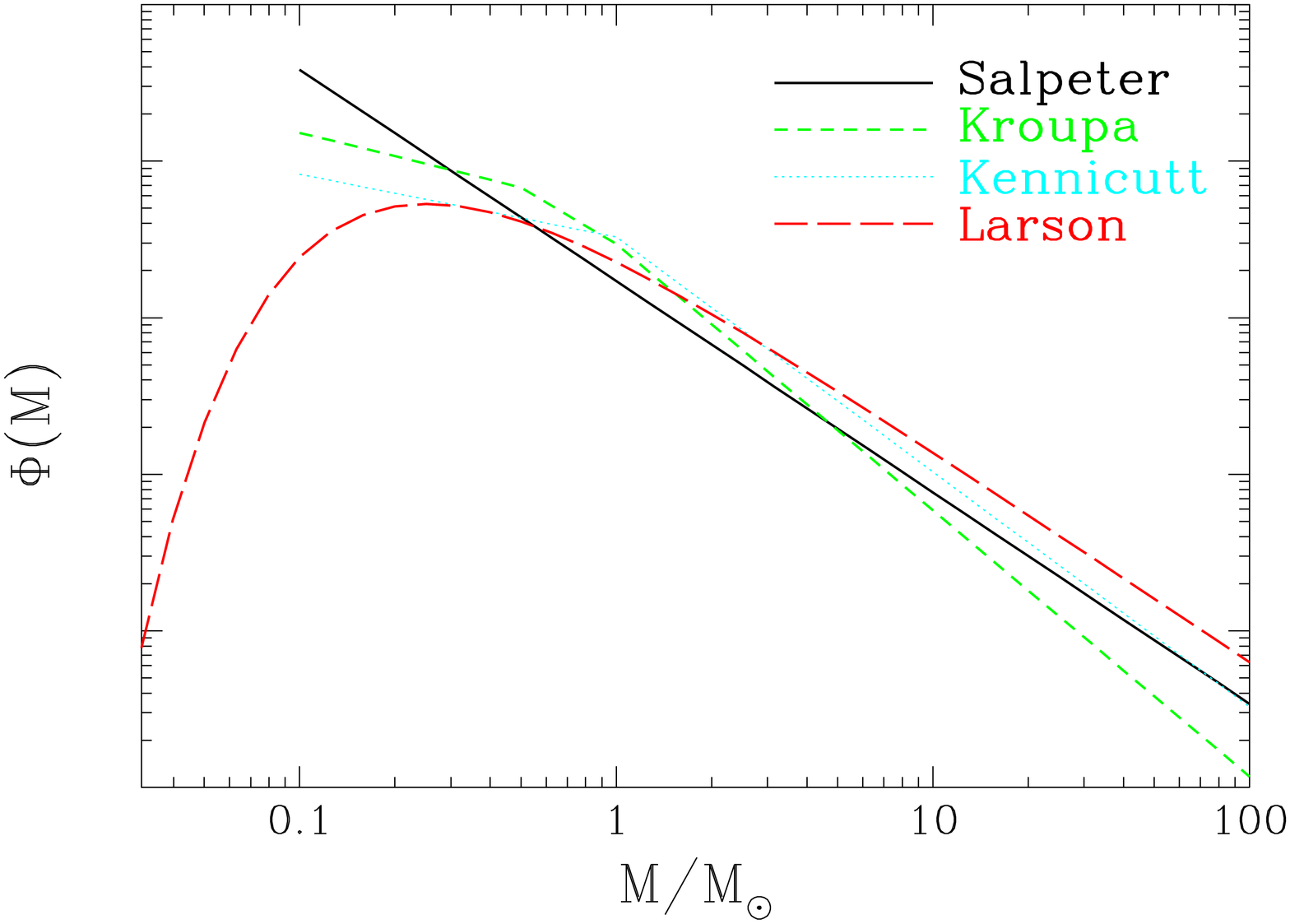,angle=270,width=5.7truecm}
\hspace{5mm}
\psfig{file=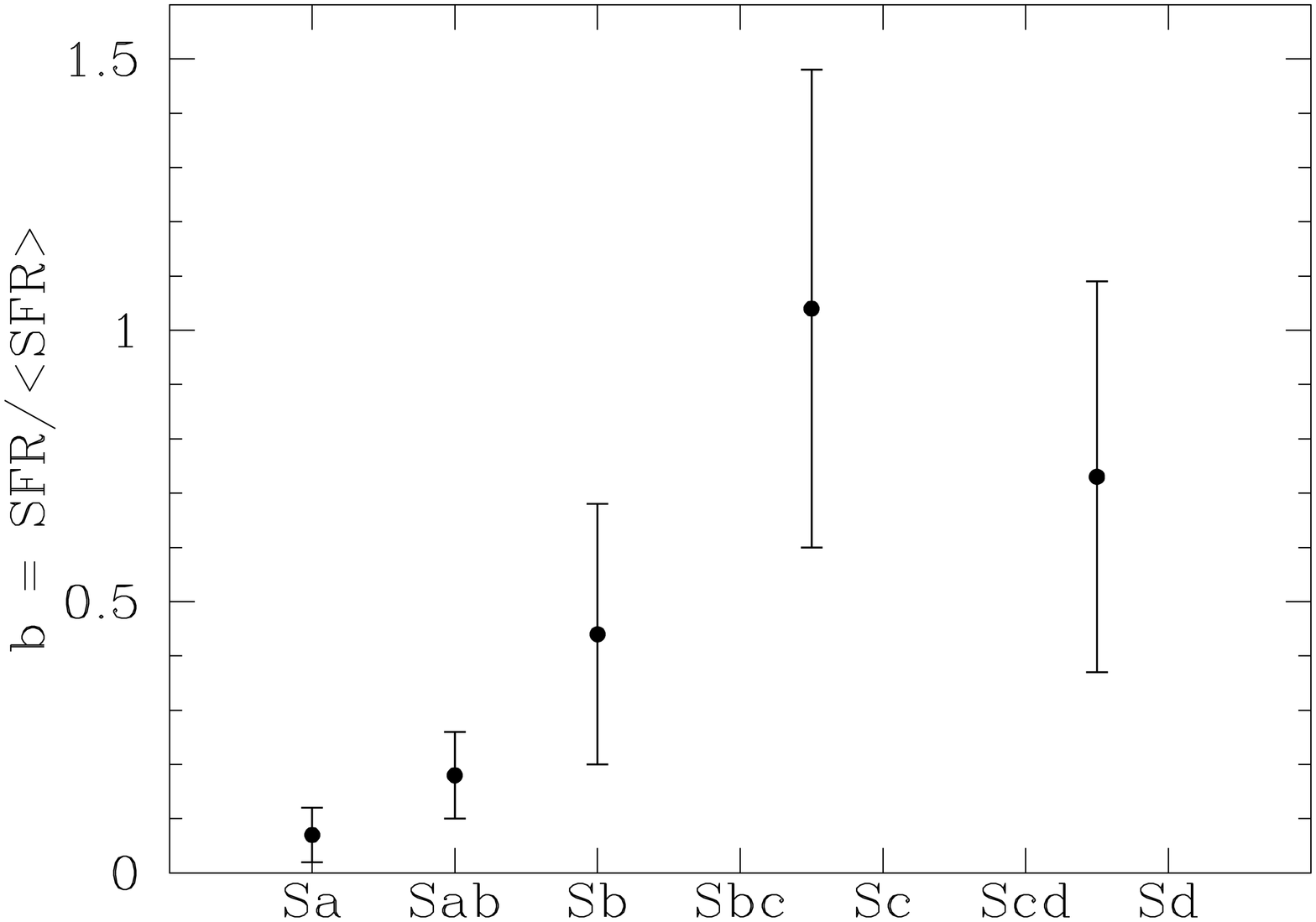,angle=270,width=5.7truecm}
\caption{{\it Left panel}: comparison between diffent IMFs, normalized
to the same total integrated mass. {\it Right panel}: mean $b$--parameter
versus Hubble type (data from Sommer--Larsen et~al.\ 2002)}
\end{figure}

Besides the IMF, the M/L ratio of a galaxy depends on its 
star formation history (SFH), that is related to Hubble type:
Kennicutt et~al.\ (1994) demonstrated that the sequence of spiral types 
is a sequence of different SFHs in the discs, as traced by the 
birthrate parameter
\[ b = \frac{SFR}{<SFR>} \]
or the ratio between the present and the past average star formation rate 
(SFR); see also Fig.~1, right panel.

The reference TF relation by Giovanelli et~al.\ (1997), suggesting the low 
M/L ratios quoted in the introduction, is typical for Sbc--Sc spirals. 
Kennicutt et~al.\ find that Sbc--Sc discs correspond to $b=0.85-1$.


\section{Results from simple models}

We will discuss in particular the M/L ratio in the I--band, 
which is an excellent optical mass tracer because
the I--band luminosity is less sensitive to recent sporadic star formation 
and to corrections for dust extinction than bluer bands. Besides,
the M/L ratio of Single Stellar Populations (SSPs) in the I--band 
is less metallicity dependent 
than in other bands, and it is less sensitive to the specific treatment 
of the AGB phase than redder, NIR bands (\S~3.1).

We computed SSPs based on the latest Padua isochrones 
(Girardi et~al.\ 2002), for the different
IMFs in \S~2. These are to be convolved with suitable SFHs.
A variety of SFHs characterized by different values of the $b$--parameter
can be simply computed adopting an exponentially decaying 
{\mbox{$SFR \propto e^{- \frac{t}{\tau}}$}} with different decaying rates.
Metallicities within a factor of 2 solar can be considered typical for
spiral galaxies. 
The resulting M/L ratio of the global stellar population
(including remnants) as a function of the {\mbox{$b$--parameter}}
is displayed in Fig.~2, for the different IMFs. 
The range $b \geq 0.8$ is representative of Sbc--Sc spirals.
The shaded area indicates the range in M/L=0.7--1 suggested in \S~1.

Fig.~2 shows that bottom--light IMFs can in fact yield M/L$_I < 1$ for 
late--type spirals, but one probably needs somewhat ``lighter''
IMFs than the local (Kroupa) one.
Full self--consistent chemo--photometric disc models are being developed 
to improve the analysis and to assess whether these IMFs are also compatible
with the observed chemical properties of spirals (Portinari et~al.\ 2002).

\begin{figure}
\begin{center}
\includegraphics[angle=270,scale=0.2]{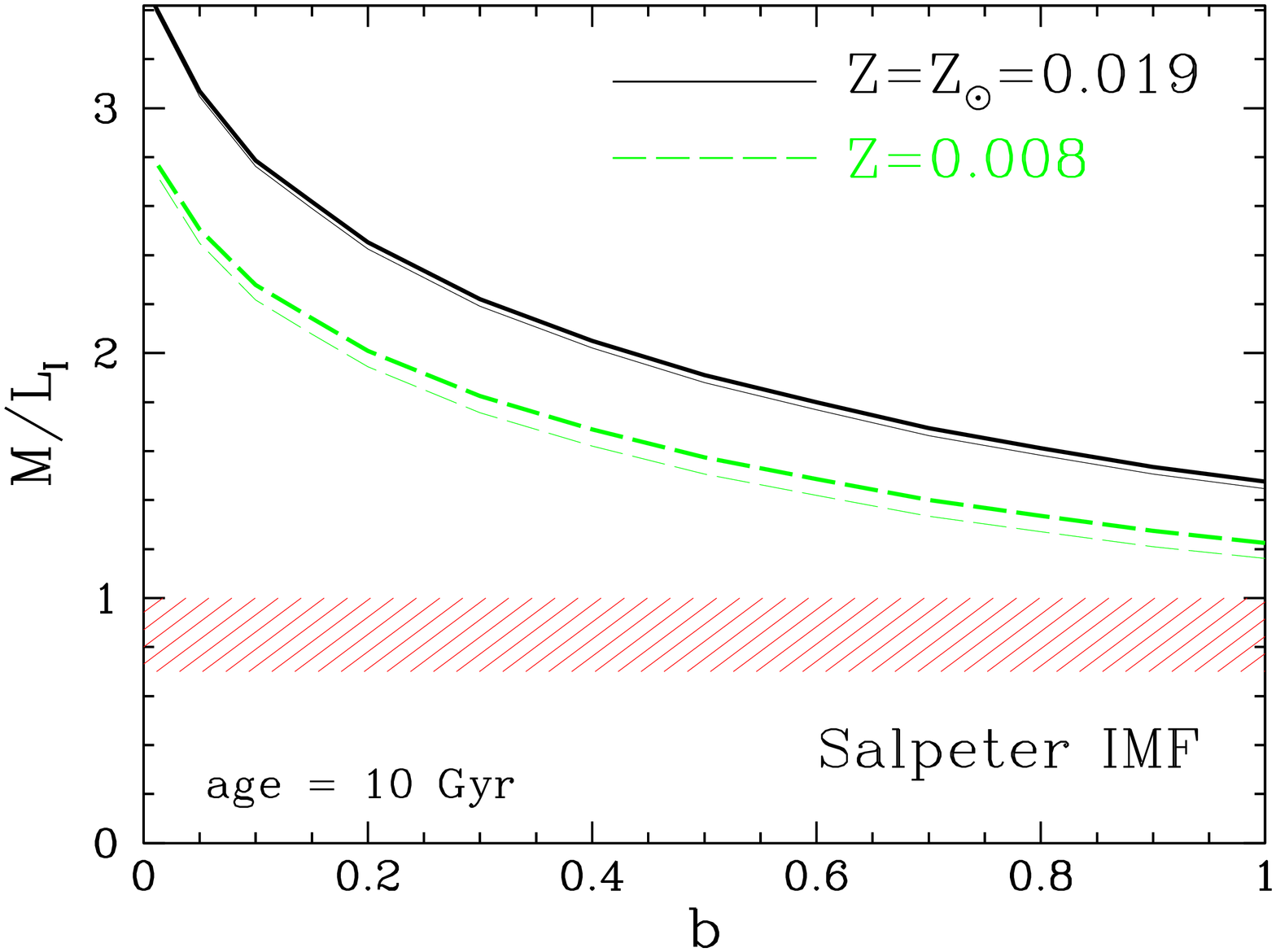}
\includegraphics[angle=270,scale=0.2]{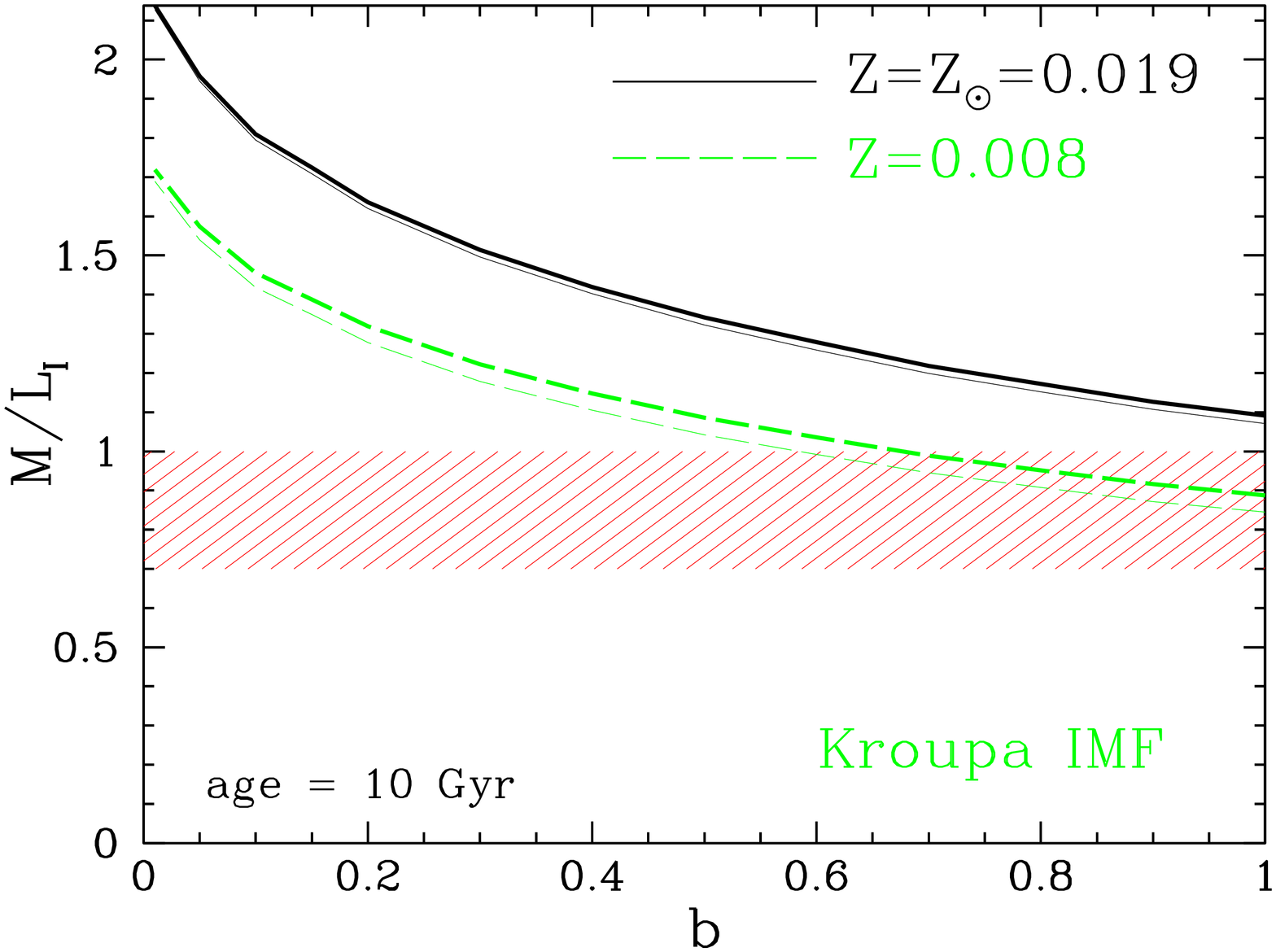}
\includegraphics[angle=270,scale=0.2]{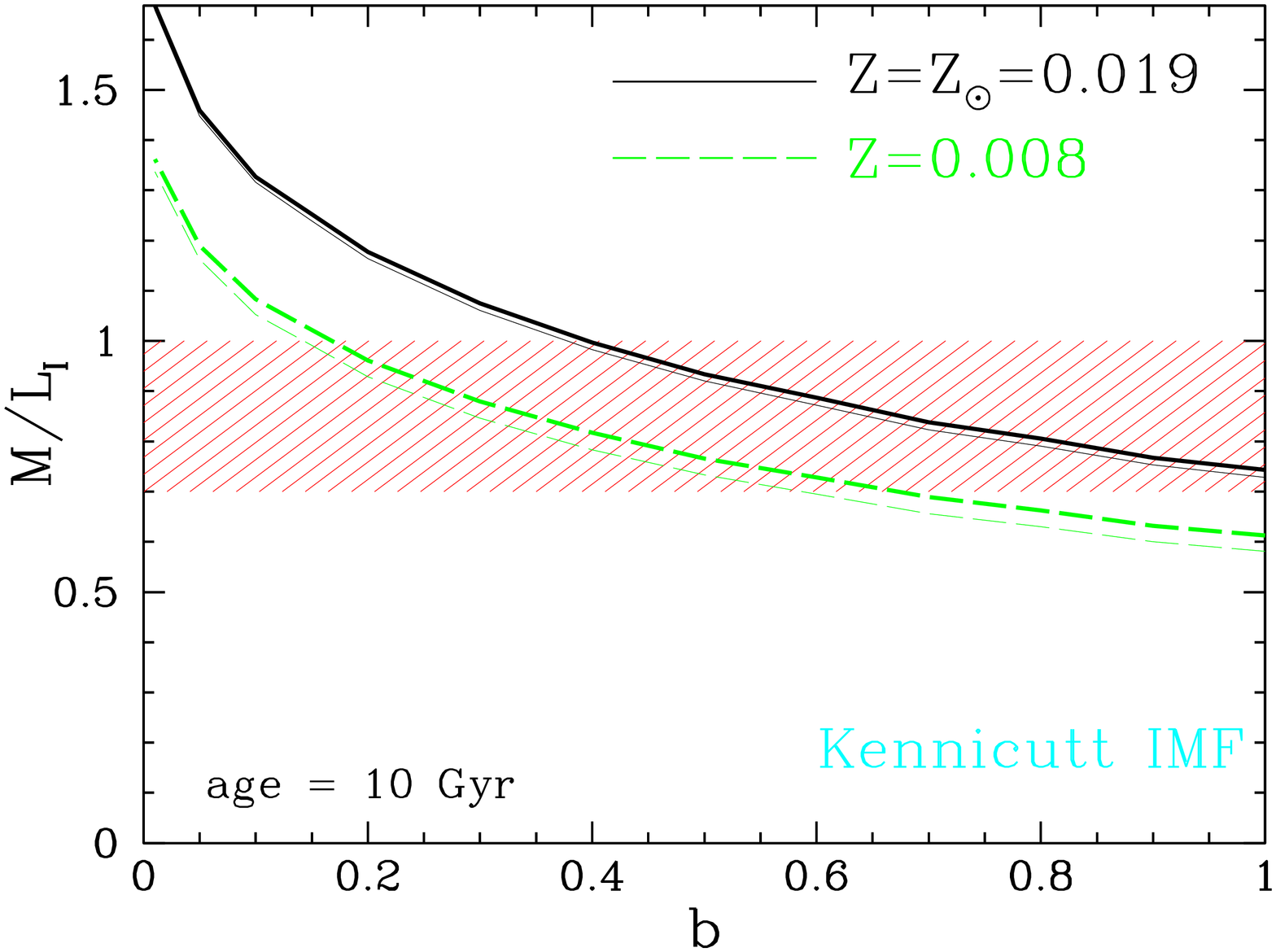}
\includegraphics[angle=270,scale=0.2]{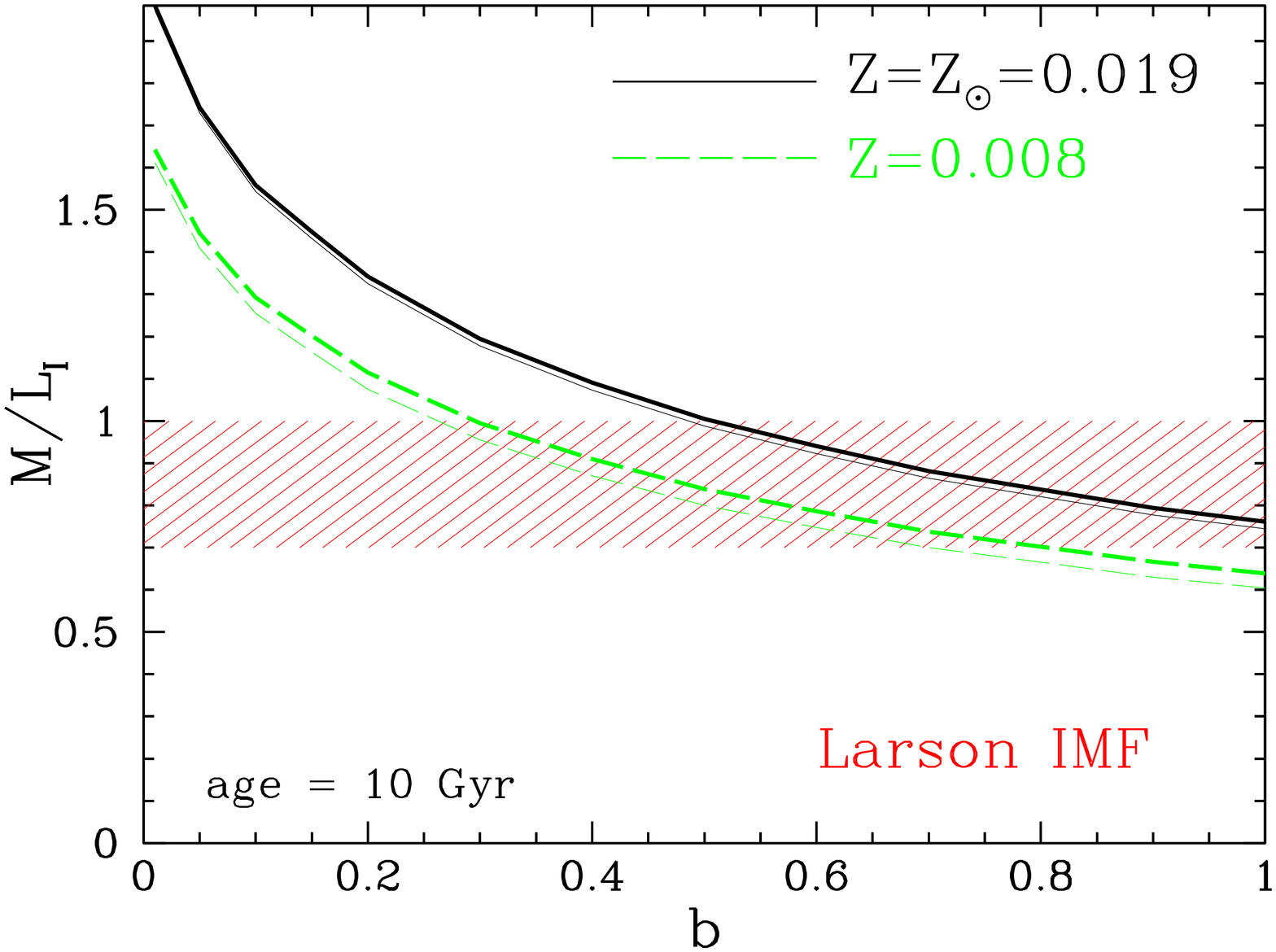}
\end{center}
\caption{M/L ratio in the I--band as a function of 
the $b$--parameter (Hubble type); $b > 0.8$ for Sbc/Sc spirals. 
Thin lines for the ``alternative SSP set''
with a more accurate treatment of the AGB (\S~4).}
\end{figure}

\vspace{-1mm}
\subsection{Effects of the AGB modelling}

\vspace{-3mm}

\noindent
We also calculated an alternative set of SSPs based on the isochrones 
by Marigo \& Girardi (2001), which incorporate a more detailed modelling 
of the Asymptotic Giant Branch (AGB) phase, with important effects on the 
integrated luminosity in the NIR bands (Fig.~3).

In Fig.~2, thin lines correspond to the alternative SSP set. The effect
on the predicted M/L ratios 
is minor in the I--band, so that results of photometric models are 
rather solid. In NIR bands like the K--band, instead, the different accuracy
in AGB
modelling can imply 20\% higher luminosities (Fig.~3, right panel), rendering
NIR luminosities a less straightforward tracer of stellar mass.

\begin{figure}
\centerline{\psfig{file=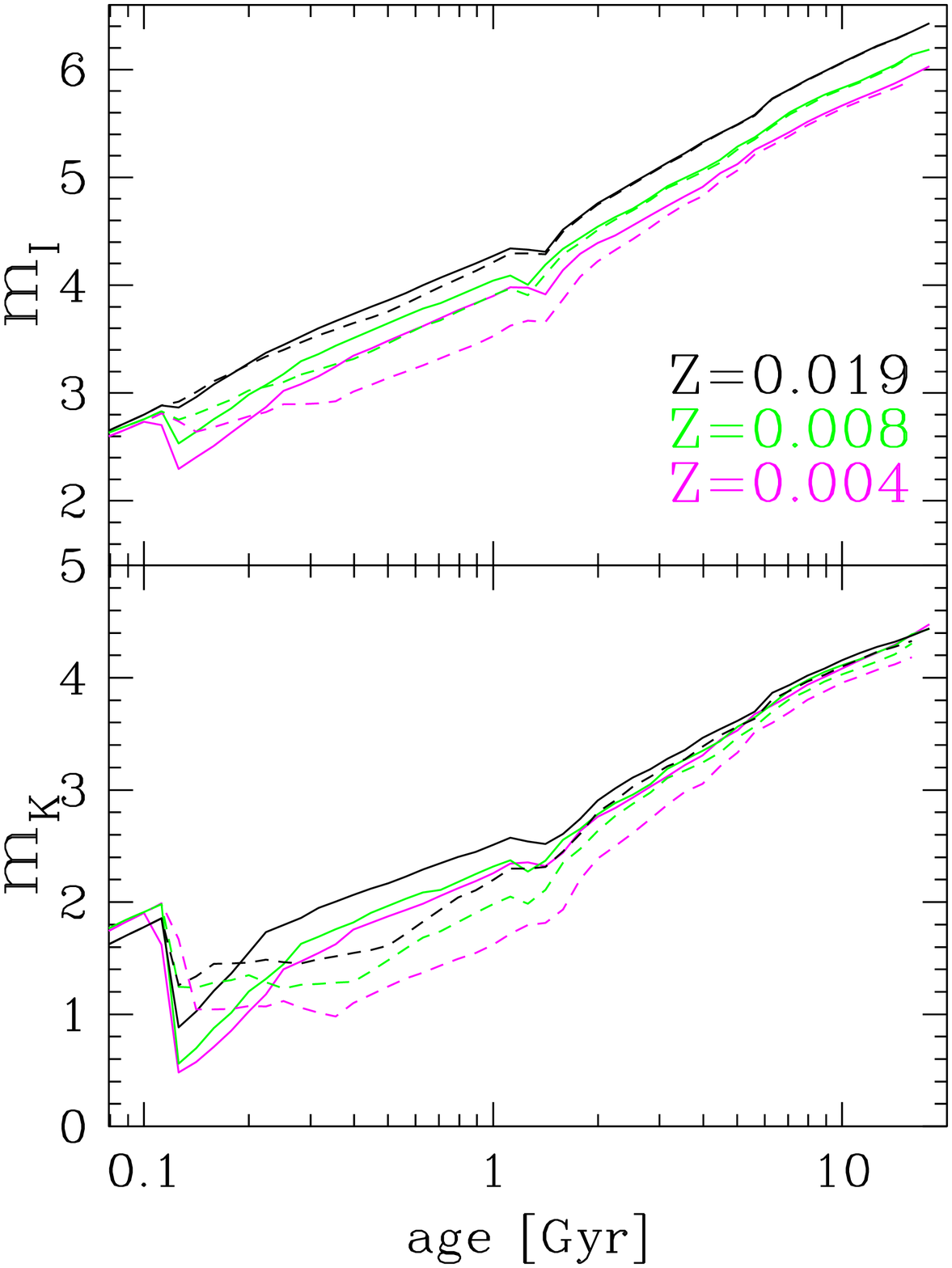,width=5truecm}
\psfig{file=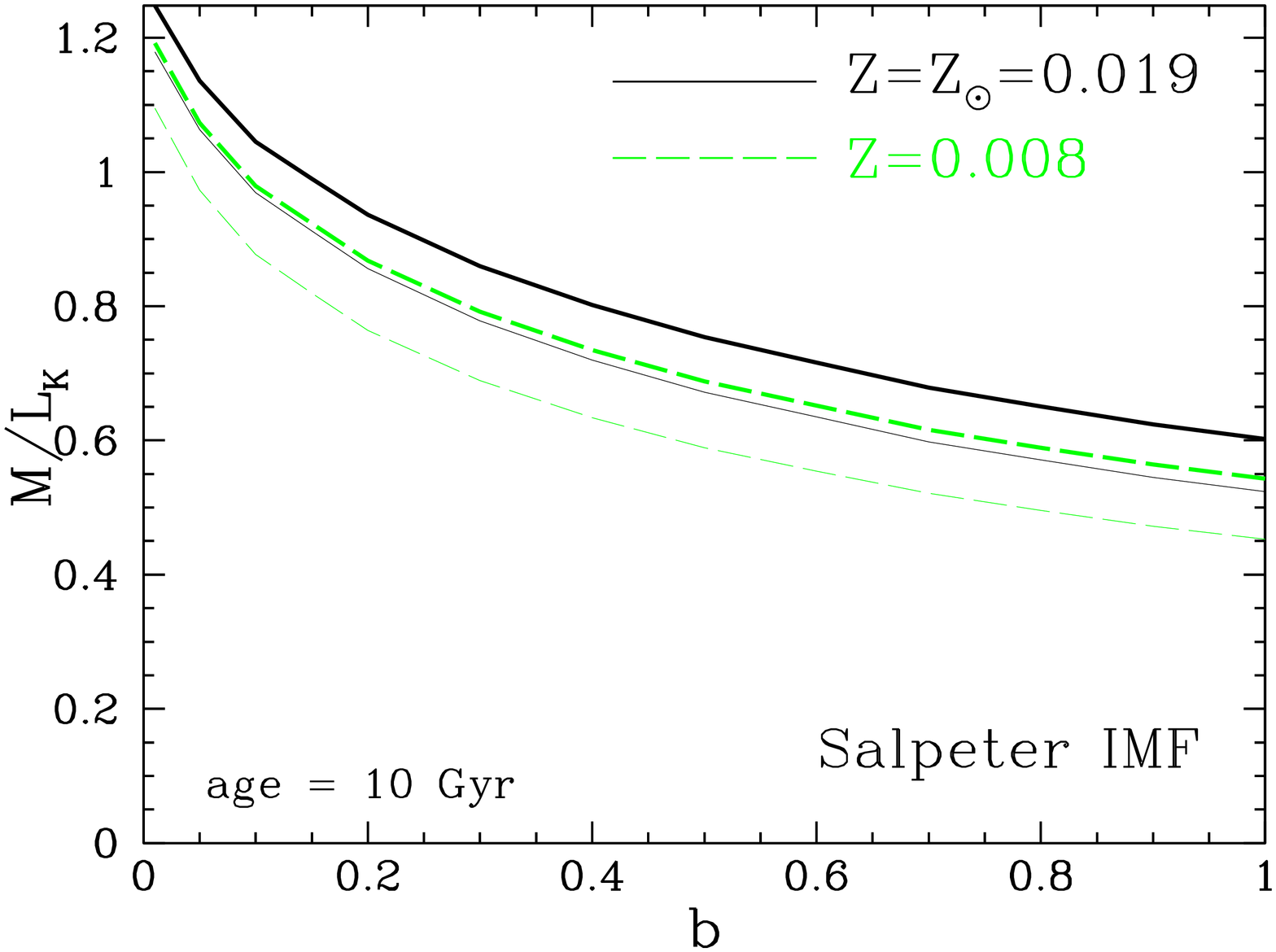,angle=270,width=6truecm} }
\caption{{\it Left panel}: luminosity evolution in the I and K bands 
for the two sets of SSPs; 
dashed lines for the case with a more accurate treatment of the AGB 
(see text). {\it Right panel}: M/L ratio in the K--band, 
comparing the two sets of SSPs; thick and thin lines as in Fig.~2.}
\end{figure}

\vspace{-5mm}


\end{article}

\end{document}